\documentclass[%
reprint,
superscriptaddress,
 amsmath,amssymb,
 aps,
pra,
]{revtex4-2}

\usepackage{graphicx}
\usepackage{graphics}
\usepackage{bm}

\begin{document}

\preprint{APS/123-QED}

\title{Two-photon superbunching effect of broadband chaotic stationary light at femtosecond timescale based on cascaded Michelson interferometer}

\author{Sheng Luo}
\email{Equal contribution}
\affiliation{MOE Key Laboratory for Nonequilibrium Synthesis and Modulation of Condensed Matter, Department of Applied Physics, Xi'an Jiaotong University, Xi'an, Shaanxi 710049, China
}
\affiliation{Electronic Materials Research Laboratory, Key Laboratory of the Ministry of Education $\&$ International Center for Dielectric Research, School of Electronic Science and Engineering, Xi'an Jiaotong University, Xi'an, Shaanxi 710049, China}%
\author{Yu Zhou}
\email{Equal contribution}
\affiliation{MOE Key Laboratory for Nonequilibrium Synthesis and Modulation of Condensed Matter, Department of Applied Physics, Xi'an Jiaotong University, Xi'an, Shaanxi 710049, China
}
\author{Huaibin Zheng}
\email{huaibinzheng@mail.xjtu.edu.cn}
\affiliation{Electronic Materials Research Laboratory, Key Laboratory of the Ministry of Education $\&$ International Center for Dielectric Research, School of Electronic Science and Engineering, Xi'an Jiaotong University, Xi'an, Shaanxi 710049, China}%
\author{Jianbin Liu}
\affiliation{Electronic Materials Research Laboratory, Key Laboratory of the Ministry of Education $\&$ International Center for Dielectric Research, School of Electronic Science and Engineering, Xi'an Jiaotong University, Xi'an, Shaanxi 710049, China}%
\author{Hui Chen}
\affiliation{Electronic Materials Research Laboratory, Key Laboratory of the Ministry of Education $\&$ International Center for Dielectric Research, School of Electronic Science and Engineering, Xi'an Jiaotong University, Xi'an, Shaanxi 710049, China}%
\author{Yuchen He}
\affiliation{Electronic Materials Research Laboratory, Key Laboratory of the Ministry of Education $\&$ International Center for Dielectric Research, School of Electronic Science and Engineering, Xi'an Jiaotong University, Xi'an, Shaanxi 710049, China}%
\author{Wanting Xu}
\affiliation{School of Science, Xi'an Polytechnic University, Xi'an, Shaanxi 710048, China}%
\author{Shuanghao Zhang}
\affiliation{Electronic Materials Research Laboratory, Key Laboratory of the Ministry of Education $\&$ International Center for Dielectric Research, School of Electronic Science and Engineering, Xi'an Jiaotong University, Xi'an, Shaanxi 710049, China}%
\author{Fuli Li}
\affiliation{MOE Key Laboratory for Nonequilibrium Synthesis and Modulation of Condensed Matter, Department of Applied Physics, Xi'an Jiaotong University, Xi'an, Shaanxi 710049, China
}
\author{Zhuo Xu}
\affiliation{Electronic Materials Research Laboratory, Key Laboratory of the Ministry of Education $\&$ International Center for Dielectric Research, School of Electronic Science and Engineering, Xi'an Jiaotong University, Xi'an, Shaanxi 710049, China}%

\date{\today}

\begin{abstract}
It is challenging for observing superbunching effect with true chaotic light, here we propose and demonstrate a method to achieve superbunching effect of $g^{(2)}(0)=2.42 \pm 0.02$ with broadband stationary chaotic light based on a cascaded Michelson interferometer (CMI), exceeding the theoretical upper limit of 2 for the two-photon bunching effect of chaotic light. The superbunching correlation peak is measured with an ultrafast two-photon absorption detector which the full width at half maximum reaches about 95 fs. Two-photon superbunching theory in a CMI is developed to interpret the effect and is in agreement with experimental results. The theory also predicts that the degree of second-order coherence can be much greater than $2$ if chaotic light propagates $N$ times in a CMI. Finally, a new type of weak signals detection setup which employs broadband chaotic light circulating in a CMI is proposed. Theoretically, it can increase the detection sensitivity of weak signals 79 times after the chaotic light circulating 100 times in the CMI.
\end{abstract}

\maketitle

\section{\label{sec:level1}introduction}
Since Hanbury Brown and Twiss (HBT) first observed the two-photon bunching effect in 1956 \cite{brown1956correlation,brown1956test,Brown1958Interferometry}, it is well known that the degree of second-order coherence (DSOC) of thermal light equals 2 for all kinds of thermal light \cite{glauber1963coherent,glauber1963photon,glauber2006nobel}, including blackbody radiation \cite{loudon1973the,tan2014measuring,Zhai2005Two}, broadband amplified spontaneous emission \cite{boitier2009measuring,hartmann2015ultrabroadband,shevchenko2017polarization,tang2018measuring}, pseudothermal light \cite{scarcelli2004two-photon,valencia2005two-photon,xiong2005experimental} and so on. Although superbunching effect of pseudothermal light has been observed \cite{hong2012two,zhou2017superbunching,bai2017photon,zhou2019experimental}, to the best of our knowledge, the superbunching effect of broadband chaotic light has never been observed. The reason is that one can not artificially generate extra intensity fluctuations in true chaotic light field like in pseudothermal light case, and also the superbunching effect of true chaotic light itself is difficult to detect due to its femtosecond timescale coherence time.

The superbunching effect of thermal light is not only academically interesting, but also has many important applications. In recent year, a serial of researches on multi-photon effect enhancement and extreme events enhancement with bright squeezed vacuum has been reported \cite{iskhakov2012superbunched,PhysRevResearch.2.013371}. The authors pointed out that these effects could be used in photon subtraction experiments \cite{manceau2019indefinite-mean}, ghost imaging \cite{Pittman1995Optical,valencia2005two-photon,Chan2009High} and  experiments which need high multi-photon effect but low average intensity \cite{spasibko2017multiphoton}.

In this paper, we proposed and demonstrated a method to achieve superbunching effect of broadband stationary chaotic light based on a cascaded Michelson interferometer (CMI), $g^{(2)}(0)=2.42 \pm 0.02$ of  stationary chaotic light is observed by using ultrafast two-photon absorption (TPA) detection technology\cite{Hayat2011Applications,nevet2011ultrafast}.  This TPA detector allows us to measure the superbuching effect of chaotic stationary light at the timescale of a few femtoseconds. The superbunching effect reported in this paper is achieved without any nonlinear process involved. A theory  based on two-photon interference is developed and explains the experimental results very well. The theory shows that comparing with ordinary bunching effect in the HBT interferometer, the superbunching effect in a CMI comes from the increasing of number of two-photon probability amplitudes involved. Also, the theoretical study shows that this superbunching effect can reach much higher value as the number of cascaded process increases, which means that the number of two-photon probability amplitudes involved increase exponentially. More importantly, the sensitivity of the CMI for small phase changes increases drastically according to our numerical simulation. Finally, we propose to use a CMI as a two-photon interferometer for weak signals detection to enhance the detection sensitivity.

The paper is organized as follows. In Sec. \ref{theory}, we will develop the quantum theory on superbunching effect of chaotic light based on two-photon interference and calculate its second-order coherence function. The method to realize superbunching effect of chaotic light is also proposed. In Sec. \ref{EXPERIMENT}, with a continuous amplified spontaneous emission (ASE) incoherent light source, a CMI and TPA detector, the superbunching effect of chaotic stationary light is observed. The discussions about the physics of superbunching of  chaotic light and its possible application are in Sec. IV.

\section{THEORY}\label{theory}
It is well known that two-photon bunching effect can be explained by two-photon interference theory at single photon level \cite{fano1961quantum}. The quantum theory of bunching effect of chaotic stationary light in the Michelson interferometer is studied in Ref. \cite{tang2018measuring}. The method to realize superbunching effect of stationary chaotic light can be found by following above studies. In the following, the quantum theory on bunching effect in the Michelson interferometer is briefly reviewed and the quantum explaination on superbunching effect of stationary chaotic light in CMI is developed.

\begin{figure*}
\includegraphics[scale=0.72]{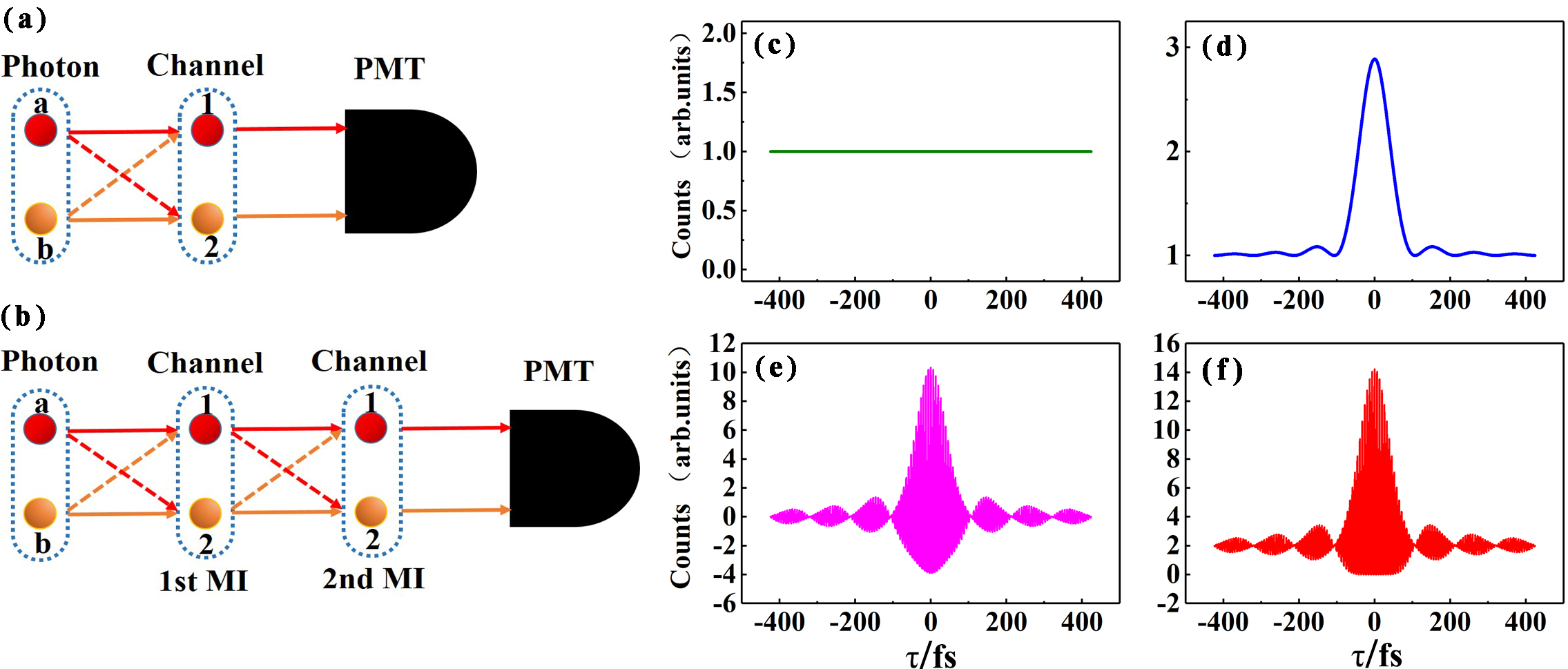}
\caption{\label{fig_1} Two-photon interference path diagram and the counts of simulated results of the chaotic light in MI. (a) A schematic diagram of the principle of two photons from chaotic light passing through MI once to trigger TPA detector. (b) A schematic diagram of the principle of two photons from chaotic light passing through the MI twice to trigger a TPA detector. The solid and dashed lines represent the photon's propagation path. PMT is a TPA detector that simultaneously receives photons from different propagation paths.  (c) shows the result of the normalization of the constant terms, corresponding to the first term in Eq. (\ref{equ5}). (d)shows the superbunching effect of chaotic light, corresponding to the sum of the second, the third and the fourth terms in Eq. (\ref{equ5}). (e) shows the sum of the high frequency oscillation terms, corresponding to the term $F(\tau)$ in Eq. (\ref{equ5}). (f) shows the sum of all the terms in Eq. (\ref{equ5}).}
\end{figure*}

The Figure \ref{fig_1}(a) shows the schematic diagram of bunching effect of chaotic light in a Michelson interferometer. We define a and b as two photons from the chaotic light. The channel 1 and channel 2 can be defined as paths that the photons are reflected from the two arms of the Michelson interferometer, respectively. There are four different but indistinguishable ways for photons $a$ and $b$ to trigger the TPA detector, corresponding to four two-photon probability amplitedes which are $A_{a 1 b 2}, A_{a 2 b 1}, A_{a 1 b 1}$ and $A_{a 2 b 2}$. The two-photon probability amplitede $A_{a 2 b 1}$ indicates that photon $a$ is reflected by mirror $2$ (in arm 2) and photon $b$ is reflected by mirror $1$ (in arm 1) before triggering the TPA detector together. The other probability amplitudes are defined similarly. In the language of quantum interference, the observed two-photon bunching effect of chaotic light in Michelson interferometer \cite{boitier2009measuring,hartmann2015ultrabroadband,shevchenko2017polarization,tang2018measuring} comes from the coherent interference of these four probability amplitudes. The probability of TPA detection $C_{TPA}$ is
\begin{equation}\label{equ1}
C_{TPA}=\left\langle|A_{a 1 b 1}+A_{a 1 b 2}+A_{a 2 b 1}+A_{a2b2}|^{2}\right\rangle.
\end{equation}

There are four different and indistinguishable paths triggering the TPA detector. In this way, the two-photon bunching effect based on the superposition of different probability amplitudes is achieved \cite{boitier2009measuring,shevchenko2017polarization,tang2018measuring}.

The superbunching effect of the broadband chaotic light can be achieved by adding more different and indistinguishable interference paths to trigger the TPA detector. In our experiment, it is realized by making chaotic light passes through a CMI. As shown in schematic diagram Fig. \ref{fig_3}(b), two photons from chaotic light pass through the same Michelson interferometer twice to trigger a TPA detector. There are 16  different and indistinguishable paths to trigger the TPA detector: $A_{a 11 b 11}, A_{a 11 b 12}, A_{a 11 b 21}, A_{a 11 b 22}, A_{a 12 b 11}, A_{a 12 b 12}, A_{a 12 b 21}\\A_{a 12 b 22}, A_{a 21 b 11}, A_{a 21 b 12}, A_{a 21 b 21}, A_{a 21 b 22}, A_{a 22 b 11}, A_{a 22 b 12}\\A_{a 22 b 21}, A_{a 22 b 22}$. The probability amplitude $A_{a 1 2 b 2 1}$ indicates that photon $a$ propagates in arm $1$ (with mirror $1$ ) when it passes the Michelson interferometer for the first time, then enters the same Michelson interferometer again and propagates in arm $2$ (with mirror $2$ ). Simultaneously photon $b$ propagates in arm $2$ and $1$ successively when it passed Michelson interferometer for the first and second time, respectively. Finally,  photon $a$ and photon $b$ trigger the TPA detection event together. The meanings of the remaining probability amplitudes are similar.

The probability of TPA detection $C_{TPA}$ is the result of the coherently superposition of the $16$ different and indistinguishable probability amplitudes, which can be expressed as \cite{book1,feynman1985qed}
\begin{equation}\label{equ2}
C_{T P A}=\left\langle\left|\begin{array}{l}
A_{a 11 b 11}+A_{a 11 b 12}+A_{a 11 b 21}+A_{a 11 b 22}+ \\
A_{a 12 b 11}+A_{a 12 b 12}+A_{a 12 b 21}+A_{a 12 b 22}+ \\
A_{a 21 b 11}+A_{a 21 b 12}+A_{a 21 b 21}+A_{a 21 b 22}+ \\
A_{a 22 b 11}+A_{a 22 b 12}+A_{a 22 b 21}+A_{a 22 b 22}
\end{array}\right|^{2}\right\rangle,
\end{equation}
where $\langle\cdots\rangle$ represents the ensemble average of all events. The probability amplitude can be written as
\begin{equation} \label{equ3}
A_{_{a m i b n j}}=e^{i \varphi_{a}} K_{a m i} e^{i \varphi_{b}} K_{b n j},
\end{equation}
where $\varphi_{a}$ and $\varphi_{b}$ are the initial phase of photon $a$ and photon $b$. $K_{a m i}$ and $K_{b n j}$ represent the Feynman propagators of different photons, where $i, j, m, n=1,2$. Since the true thermal light source has a certain spectral distribution, which is different from the single frequency of the laser. In this case, the Feynman propagator from the thermal light can be expressed as \cite{peskin1995an}
\begin{equation}\label{equ4}
K\left(r_{2}, t_{2} ; r_{1}, t_{1}\right)=\int_{\omega_{0}-\frac{1}{2} \Delta \omega}^{\omega_{0}+\frac{1}{2} \Delta \omega} f(\omega) e^{-i\left[\sigma\left(t_{1}-t_{2}\right)-\bar{k}_{12} \cdot \bar{r}_{12}\right]} d \omega,
\end{equation}
where $f(\omega)$ is the spectral distribution function of light source that assumed the rectangular spectrum distribution. $\omega_{0}$ is the center frequency and $\Delta\omega$ is the spectral bandwidth. All these parameters depend on the spectral characteristics of the light source we used. $t_{1}$, $t_{2}$ represent the time at which the photon arrives at $r_{1}$ and $r_{2}$ respectively, so the $t_{1}-t_{2}$ is the time difference between the photon from $r_{1}$ and $r_{2}$. $\vec{k}_{12}$ represents the vector of light waves where the photons are from $r_{1}$ to $r_{2}$, and $\vec{r}_{12}$ represents the difference between the position vectors of the photons from $r_{1}$ to $r_{2}$.

Here we only concentrate on the temporal correlation of thermal light for simplicity. So substituting Eq. (\ref{equ3}) and Eq. (\ref{equ4}) into Eq. (\ref{equ2}) can get
\begin{eqnarray}\label{equ5}
\begin{aligned}
C_{T P A} &=18(\Delta \omega)^{2}+18(\Delta \omega)^{2}+2(\Delta \omega)^{2} \operatorname{sinc}^{2}\left(\Delta\omega\tau\right)\\
&+32(\Delta\omega)^{2}\operatorname{sinc}^{2}\left(\frac{1}{2}\Delta\omega\tau\right)+F(\tau).
\end{aligned}
\end{eqnarray}

The normalized simulation results are shown in Fig. \ref{fig_1} which the background of the constant term is 1. $18(\Delta \omega)^{2}$ is the constant term corresponds to the Fig. \ref{fig_1}(c), $18(\Delta \omega)^{2}+2(\Delta \omega)^{2} \operatorname{sinc}^{2}\left(\Delta \omega\tau\right)+32(\Delta \omega)^{2}\operatorname{sinc}^{2}\left(\frac{1}{2} \Delta \omega\tau\right)$ is the superbunching interference term as shown in the Fig. \ref{fig_1}(d). $F(\tau)$ is the high-frequency oscillation term which is present in Fig. \ref{fig_1}(e), where $F(\tau)=32(\Delta\omega)^{2}\cos\left(2\omega_{0}\tau\right)\operatorname{sinc}^{2}\left(\frac{1}{2} \Delta \omega\tau\right)+2(\Delta \omega)^{2}\cos \left(4 \omega_{0}\tau\right)\cdot\operatorname{sinc}^{2}\left(\Delta \omega\tau\right)+96(\Delta \omega)^{2} \cos \left(\omega_{0}\tau\right) \operatorname{sinc}\left(\frac{1}{2} \Delta \omega\tau\right)+24(\Delta \omega)^{2} \cdot\\ \cos\left(2 \omega_{0}\tau\right) \operatorname{sinc}\left(\Delta \omega\tau\right)+32(\Delta \omega)^{2} \cos \left(\omega_{0}\tau\right) \operatorname{sinc}\left(\frac{1}{2} \Delta \omega\tau\right)\cdot \cos \left(2 \omega_{0}\tau\right) \operatorname{sinc}\left(\Delta \omega\tau\right)$. The result of Fig. \ref{fig_1}(f) signifies the sum of all the terms corresponds to the whole detection event $C_{T P A}$. The normalizing the second-order coherence function is given by
\begin{equation}\label{equ6}
\begin{aligned}
g^{(2)}\left(\tau\right)=1+\frac{16}{9} \operatorname{sinc}^{2}\left(\frac{1}{2} \Delta \omega\tau\right)+ \frac{1}{9} \operatorname{sinc}^{2}\left(\Delta \omega\tau\right),
\end{aligned}
\end{equation}
where $\operatorname{sinc}(x)=\sin (x)/x$. When $\tau$ equals zero, $g^{(2)}\left(\tau\right)=2.89$, which means the superbunching effect of the broadband chaotic light can be observed in the CMI scheme. 

\section{EXPERIMENTAL IMPLEMENTATION AND RESULTS }\label{EXPERIMENT}

The method of observing superbunching effect of chaotic stationary light is to make the chaotic light pass through a CMI. In the CMI the two-photon interference effect is enhanced and the superbunching effect can be observed. The detailed theoretical analysis is in Sec. II and in this section we will focus on how to realize the CMI and report the experimental results.

\begin{figure}[htbp]
\flushleft
\includegraphics[scale=0.3]{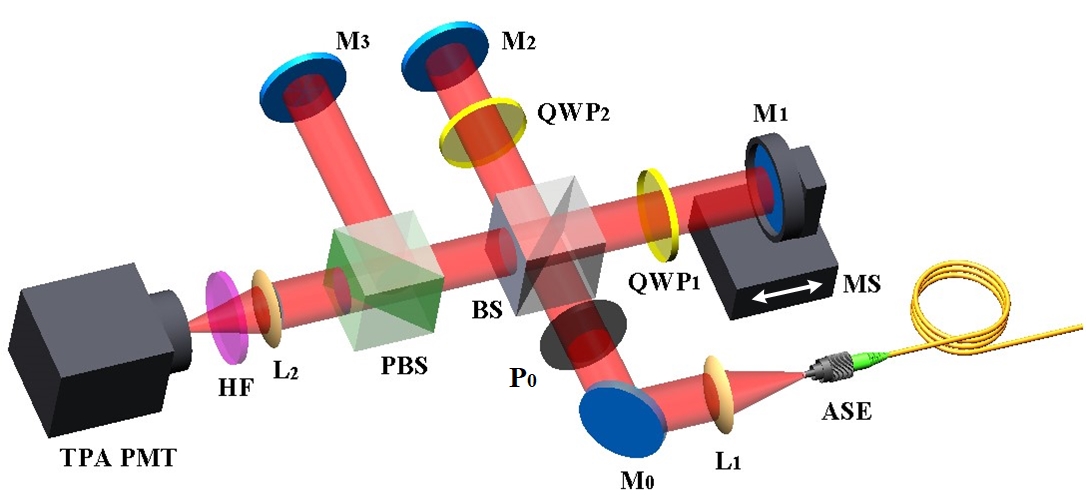}
\caption{\label{fig_2} Experimental setup of superbunching thermal light based on CMI. $\rm M_{0}$, $\rm M_{1}$, $\rm M_{2}$ and $\rm M_{3}$ are 1550nm dielectric mirrors, where $\rm M_{1}$ is fixed on a precise motorized linear translation stage (MS). PMT is a GaAs photomultiplier tube.}
\end{figure}

The experimental setup for observing superbunching effect of broadband chaotic light is shown in Fig. \ref{fig_2}. The light source is a continuous amplified spontaneous emission (ASE) incoherent light. Its center wavelength is 1550 nm with bandwidth of 30 nm. $\rm L_{1}$ and $\rm L_{2}$ are two convergent lenses with focal lengths of 10 mm and 25.4 mm respectively. $\rm P_{0}$ is a linear polarizer with horizontal direction. $\rm QWP_{1}$ and $\rm QWP_{2}$ are two quarter-wave plates with an angle of $45^{\circ}$ between the optical axes and the polarization direction of the incident light. The broadband light is collimated by $\rm L_{1}$ and passes through $P_{0}$, which causes the unpolarized chaotic light to be horizontally polarized and  enters the Michelson interferometer for the first time. After through the non-polarizing beam-splitter cube (BS), two split light beams pass through $\rm QWP_{1}$ and $\rm QWP_{2}$, then are reflected back from $\rm M_{1}$ and $\rm M_{2}$, then pass through $\rm QWP_1$ and $\rm QWP_2$ again respectively. At this moment  the polarizations of two beams changed from horizontal to vertical. Then all vertically polarized light will be reflected to mirror $\rm M_{3}$ by a polarizing beam-splitter cube (PBS). The light reflected by mirror $\rm M_{3}$ will  go into the Michelson interferometer for the second time. By manipulating the polarization of light beams, we realize the function of a CMI. Finally, after the light interferes in the Michelson interferometer the second time the  vertically polarized light becomes  horizontally polarized because of $\rm QWP_{1}$ and $\rm QWP_{2}$. Then two horizontally polarized light beams reflected from mirrors $\rm M_{1}$ and $\rm M_{2}$ pass through the PBS, a spherical lens ($\rm L_{2}$) and a high pass filter (HF) to trigger a H7421-50 Hamamatsu GaAs-photomultiplier tube (PMT) operated in two-photon absorption mode \cite{boitier2009measuring}. During the experiment the dark counts is 80 per second. The spherical lens and the high-pass filter with cutoff frequency of 1300nm is used to eliminate the single-photon counting of the PMT and make sure the PMT working in TPA mode.

\begin{figure}[htbp]
\flushleft
\includegraphics[scale=0.61]{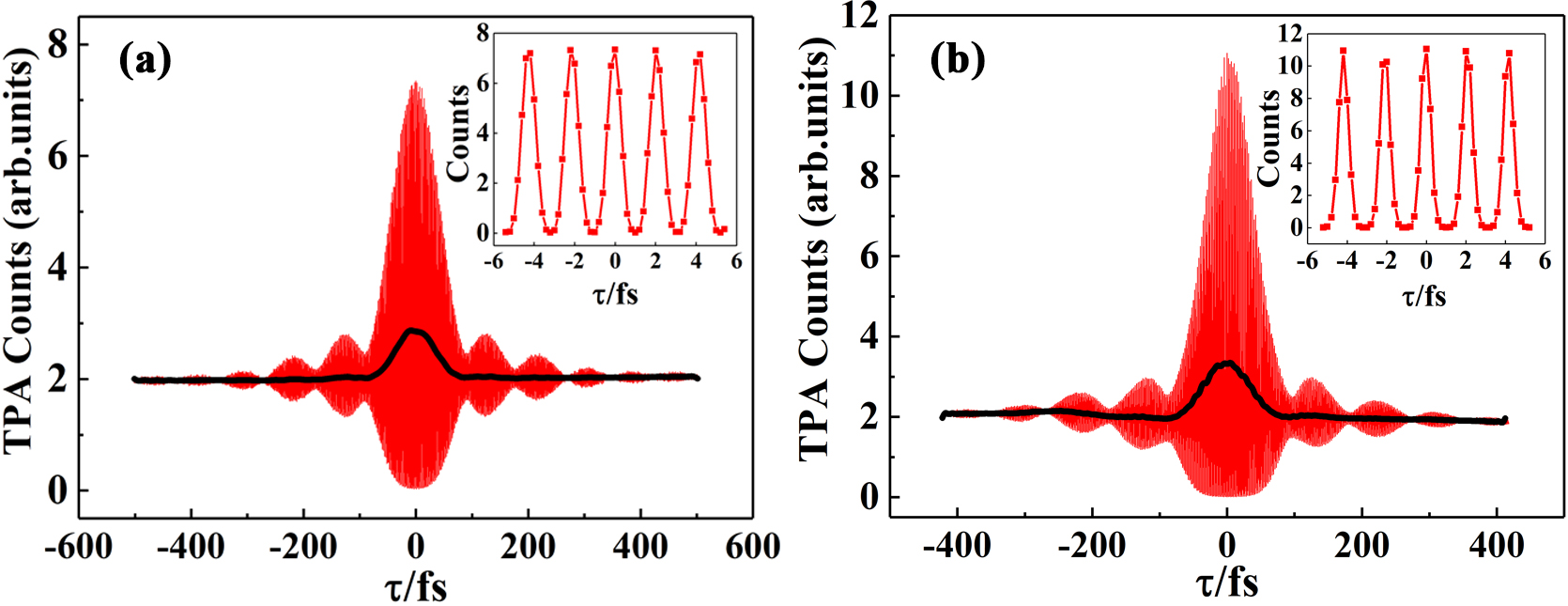}
\caption{\label{fig_3} (a) The measurement results of the broadband chaotic light bunching effect when light passes through the Michelson interferometer. (b) The measurement results based on the broadband chaotic light superbunching effect when light passes through the Michelson interferometer twice in succession. The red line is a graph of all counting results of the TPA detector. The black solid line is the result of filtering out high-frequency terms after numerical signal processing. The insets in Fig. \ref{fig_3} (a) and (b) illustrate enlarged detail of the interferogram near 0 fs, respectively.}
\end{figure}

To make a comparison between ordinary bunching effect \cite{boitier2009measuring,hartmann2015ultrabroadband,shevchenko2017polarization} and superbunching effect of broadband stationary chaotic light at femtosecond timescale, we first measured its photon bunching effect. This is realized by removing the $\rm PBS$ and mirror $\rm M_{3}$ from the setup shown in Fig. \ref{fig_2}. Without the PBS and mirror $\rm M_{3}$ the setup is a standard Michelson interferometer and chaotic light in it circulate only once before been detected by the TPA detector. The second-order correlation function of bunching effect is measured by scanning one arm of the interferometer--scanning mirror $\rm M_{1}$ on a mortor stage (MS). The TPA counts is recorded versus the time delay difference between two arms. The experimental results are shown in Fig. \ref{fig_3}(a). The red line is the measured second-order correlation function which full width at half maximum (FWHM) is about 123 fs, and the visibility of the second-order interference pattern can reach to 99.2\%. The black solid line is the result of filtering out the high-frequency oscillation terms, and its peak-to-background ratio is about 1.44, which corresponds to the second-order coherence function $g^{(2)}(0)=1.87 \pm 0.02$ is obtained after normalization. The oscillating of TPA counts near the maximum of correlation peak are shown in greater detail in the inset of Fig. \ref{fig_3}. The measurement can be understood using Refs. \cite{boitier2009measuring,tang2018measuring}, these researches greatly facilitated the development of measuring the bunching effect based on chaotic stationary light.

In the next step, the $\rm PBS$ and mirror $\rm M_{3}$ are put back into the setup to constitute the CMI as shown in Fig. \ref{fig_2}. In this setup, chaotic light will circulate twice in the CMI before detected by the TPA detector. By doing so, the two-photon interference paths increase and the two-photon interference effect is enhanced.  The experiment results are shown in Fig. \ref{fig_3}(b). The red line is the measured second-order correlation function which FWHM is about 95 fs and the visibility of the second-order interference pattern can reach to 99.8\%. The black solid line is the result of removing the high-frequency oscillation term by numerical filtering and its peak-to-background ratio is about 1.71,  which corresponds to the second-order coherence function $g^{(2)}(0)=2.42 \pm 0.03$ is obtained after normalization. The oscillating of TPA counts near the maximum of correlation peak are shown in greater detail in the inset of Fig. \ref{fig_2}(b).

Comparing the results of Fig. \ref{fig_3}(a) and Fig. \ref{fig_3}(b) we can see that the  the superbunching effect of broadband stationary chaotic light was observed in a CMI at femtosecond timescale .

\section{DISCUSSIONS}\label{discussion}
\subsection{$N$-order superbunching effect of chaotic light}\label{discussion1}
Based on the above work, it is easy to think that the increasing number of propagating through Michelson interferometer is actually adding different and indistinguishable two-photon interference paths, it is the reason that we observed superbunching effect of the broadband chaotic light in the CMI. When the beam passes $N$ times through Michelson interferometer, the all TPA detection event can be expressed as
\begin{equation}\label{equ7}
C_{T P A}=\left\langle\left|\sum_{i, j=1,2}^{4^{N}} A_{a i_{1} i_{2} \ldots i_{n}, b j_{1} j_{2}\ldots j_{n}}\right|^{2}\right\rangle,
\end{equation}
where $A_{a i_{1} i_{2} \ldots i_{n}, b j_{1} j_{2} \ldots j_{n}}$ is the probability amplitude of photons $a$ and $b$ passing through $N$ times the arms of the Michelson interferometer respectively, $i_{1}, i_{2} \cdots, i_{n}, j_{1}, j_{2} \cdots, j_{n}=1,2$.

However, it is too difficult to calculate each term of all events, mainly because the probability amplitude has increased to $4^{\mathrm{N}}$ as many terms as possible. Therefore, for the calculation of DSOC after transferring $N$ times through Michelson interferometer, a new statistical induction method was proposed based on the calculation of the coefficient of the constant $C_{c}$ and second-order coherence terms $C_{i}$. As the number of times passing through Michelson interferometer is increased, the high-frequency oscillation terms that required to filter out will be more and more, and it will not make any contributions to the calculation of the second-order coherence function. Therefore, we only calculate the coefficient of $C_{c}$ and $C_{i}$ which are straightforwardly related to the second-order coherence function.

After derivation, when the chaotic light transfers $N$ times through Michelson interferometer, the coefficient of constant terms $C_{c}$ can be expressed in the TPA detection event as
\begin{equation}\label{equ8}
C_{c}=\left(\sum_{n=0}^{N}\left(C_{N}^{n}\right)^{2}\right)^{2},
\end{equation}
similarly, the coefficient of second-order coherence terms $C_{i}$ is statistically calculated as
\begin{equation}\label{equ9}
C_{i}=\sum_{m=1}^{N} 2 \cdot\left(\sum_{n=0}^{N-m} C_{N}^{n} C_{N}^{m+n}\right)^{2}.
\end{equation}

The specific details of $C_{c}$ and $C_{i}$ can be found in Appendix \ref{Appendix A} and \ref{Appendix B}. Where $C_{N}^{n}$ indicates that combinations of $n$ are taken from $N$ different elements, $N \geq n$. The meaning of $C_{N}^{m+n}$ is the same.
\begin{figure}[htbp]
\includegraphics[scale=0.29]{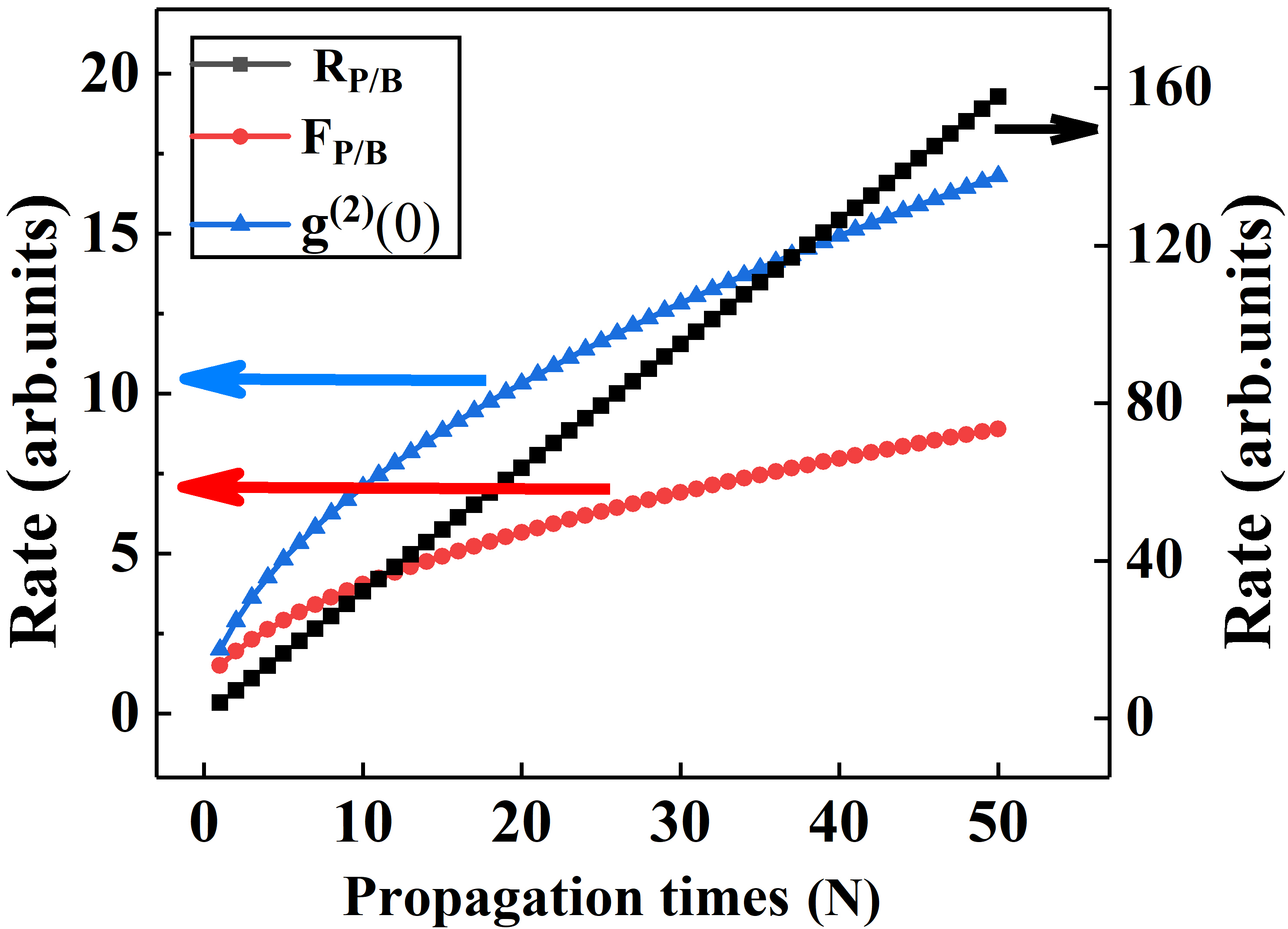}
\caption{\label{fig_4} Ratio of two-photon interference pattern of thermal light after transferring $N$ times through the Michelson interferometer under different conditions. The black rectangle represents $R_{P / B}$. The red circle represents $F_{P / B}$. The blue triangle represents  $g^{(2)}(0)$, that is the degree of the second-order coherence of the two-photon superbunching.}
\end{figure}

In order to further study the $N$-order chaotic light superbunching effect, $C_{c}$ and  $C_{i}$ are derived in the Eq. (\ref{equ8}) and Eq. (\ref{equ9}). When the beam passes through Michelson interferometer $N$ times, first we can get that the peak-to-background ratio of the interference pattern of the whole TPA detection is
\begin{equation}\label{equ10}
R_{P / B}=\frac{16^{N}}{C_{c}},
\end{equation}
then the peak-to-background ratio of the two-photon interference pattern after filtering out the high-frequency oscillation terms shown as
\begin{equation}\label{equ11}
F_{P / B}=1+\frac{C_{i}}{C_{c}}.
\end{equation}
Naturally, expression of the degree of the second-order coherence which we concerned about is
\begin{equation}\label{equ12}
g^{(2)}(0)=1+2 \frac{C_{i}}{C_{c}}.
\end{equation}

The simulations of Eqs. (\ref{equ10}), (\ref{equ11}) and (\ref{equ12}) are shown in Fig. \ref{fig_4}, where $R_{P / B}$ increases linearly, $F_{P / B}$ and $g^{(2)}(0)$ increase logarithmically with the increase of $N$. When $N$ = 50, $F_{P / B}=8.90 \pm 0.01$, $R_{P / B}=157.87 \pm 0.02$ and $g^{(2)}(0) =16.79 \pm 0.01$ can be obtained. Moreover, it is obviously that the increasing trend of $g^{(2)}(0)$ will gradually slow down with the increase of propagation times $N$. Principally as the number of times passing through the Michelson interferometer increases, the first-order interference terms inside will increase exponentially. Therefore, the increasing oscillation terms will lower the visibility and slow down DSOC growth. Nonetheless, it does not prevent us from achieving the superbunching effect of chaotic light.

\subsection{A new type of weak signals detection setup based on multi-photon interference}\label{discussion2}
\begin{figure}[htbp]
\includegraphics[scale=0.32]{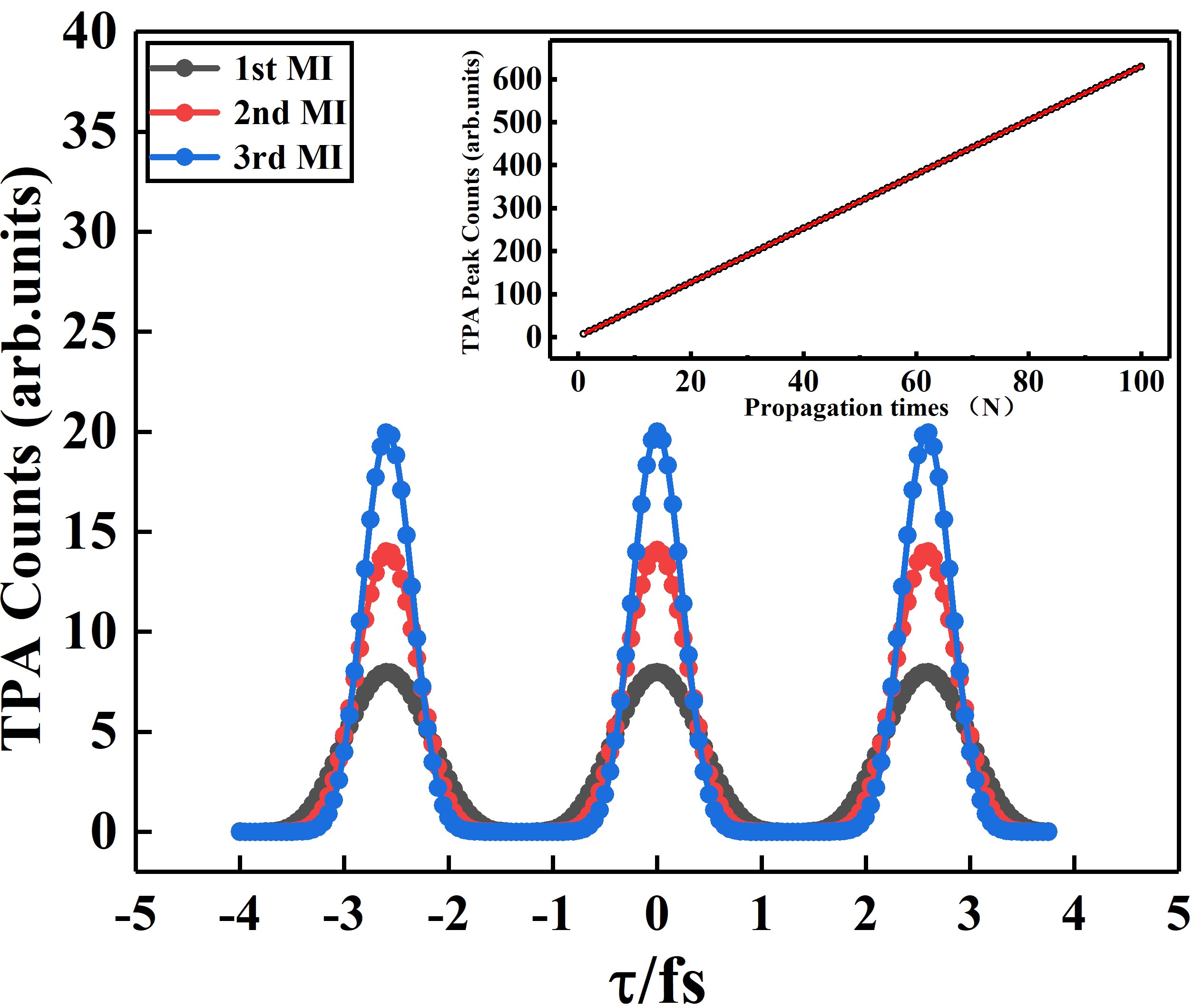}
\caption{\label{fig_5} The oscillatory behaviors of TPA interferogram under different propagation times. The black, red and blue circle represents details of TPA detection event when chaotic light transfers through the CMI once, twice and thrice, respectively. The inset in Fig. \ref{fig_5}
represents the trend of the TPA peak counts as propagation time increases.}
\end{figure}

From above studying, we notice that the superbunching effect in the CMI could enhance the sensitivity of phase detection. This is because multiple circulation of chaotic light in the CMI leads to multi-photon interference. Hence we proposed to employ chaotic light circulating in the CMI to increasing the detection sensitivity for weak signals detection. The insets in Fig. \ref{fig_3} (a) and (b) illustrate enlarged detail of the interferogram near 0 fs, respectively. It is obvious that the peak counts of chaotic light circulating in the CMI twice are higher than those in the cascade MI once, which are 11.05 and 7.34, respectively. The numerical simulation is shown in Fig. \ref{fig_5} corresponding to the measurement results of insets of Fig. \ref{fig_3}. The black, red and blue circle represents details of TPA detection event when chaotic light transferring through the CMI once, twice and thrice, respectively. When $\tau$ equals zero, the theoretical TPA peak counts are 8, 14.22 and 20.48 separately and the period of the interference pattern is 2.55 fs. The specific theoretical derivation of chaotic light transferring through the CMI thrice is similar to Sec. \ref{theory} above. It is assumed that the relative TPA counts will change from 0 to 8, 14.22 and 20.48 separately when the phase changes from 0 to $\pi$. The TPA peak count increases with the increase of propagation time, which means the weak signal is amplified under the same phase variation.

Based on the derivation of $C_{c}$ and $C_{i}$, we found that the meaning of TPA peak counts is the same as Eq. (\ref{equ9}) besides the different background. Therefore, the TPA peak counts can be expressed as $2R_{P / B}$. The inset in Fig. \ref{fig_5} illustrates TPA peak counts linearly increases with the increase of propagation time. When $N$ = 100, TPA peak counts equals 629.89, which means the signal amplifies 79 times compared with chaotic light circulating in the CMI once. The enhancement effect is expected to increase with the light circulating more times in the CMI. It should be emphasised that enhancement effect is the result of multi-photon interference, as the number of times passing through the Michelson interferometer increases, the interference terms will increase exponentially. In our design, the circulation of chaotic light in the CMI provides more and more two-photon probability amplitudes interfering with each other, it not only realize the superbunching effect of the broadband chaotic light, but also significantly increases detection sensitivity, which provides a new solution for the detection of weak signals.

\section{CONCLUSIONS}
In summary, two-photon superbunching effect of broadband chaotic light is proposed and demonstrated at femtosecond timescale based on the CMI. DSOC of chaotic light was measured as $2.42 \pm 0.03$. Two-photon interference theory, which describes superbunching effect of the broadband chaotic light in a cascade MI, has also been developed. Furthermore, we demonstrated the expression of superbunching chaotic light by deriving the coefficients of the constant and second-order coherence terms creatively when beams passing through Michelson interferometer $N$ times, whose variational trends will raise with increasing $N$. Also our numerical simulation shows that the N-order cascaded Michelson interferometer dramatically improve sensitivity of detecting small phase changes. A new type of weak signals detection setup could be designed based on the superbunching effect reported in the paper. As this kind of superbunching chaotic light is simple to realize and control, it will lay a good foundation for future research on two-photon interferometer for weak signals detection.
\begin{acknowledgments}
This work was supported by Shaanxi Key Research and Development Project (Grant No. 2019ZDLGY09-09); National Natural Science Foundation of China (Grant No. 61901353); Key Innovation Team of Shaanxi Province (Grant No. 2018TD-024) and 111 Project of China (Grant No.B14040).
\end{acknowledgments}

\appendix
\section{The coefficient of the constant terms $C_{c}$ of chaotic light transferring through the Michelson interferometer $N$ times}\label{Appendix A}
When the chaotic light transfers through the Michelson interferometer (MI) $N$ times, there are $4^{N}$ kinds of different and indistinguishable two-photon interference paths, the whole two-photon absorption detection event can be expressed as Eqs. (\ref{equ6}).

There will be $16^{N}$ terms after modulus square is calculated. Any one of them can be expressed as $A_{a i_{1} i_{2} \cdots i_{n}, b j_{1} j_{2} \cdots j_{n}} A_{a p_{1} p_{2} \cdots p_{n}, b q_{1} q_{2} \cdots q_{n}}^{*}$, where $i_{1}, i_{2}, \cdots i_{n}, j_{1}, j_{2} \cdots j_{n}=1,2$ ,  $p_{1}, p_{2} \cdots p_{n}, q_{1}, q_{2} \cdots q_{n}=1,2$, $A_{a p_{1} p_{2} \cdots p_{n}, b q_{1} q_{2} \cdots q_{n}}^{*}$ is the complex conjugate of $A_{a i_{1} i_{2} \cdots i_{n}, b j_{1} j_{2} \cdots j_{n}}$, $a$ and $b$ represent two photons from the chaotic light, respectively. These symbols indicate whether each photon travels through channel 1 or channel 2 through the MI. For example, $i_{n}$ indicates that the photon passes through the channel i when it passes through the $N$-time MI.

Here we take  $A_{a i_{1} i_{2} \cdots i_{n}, b_{1} j_{2} \cdots j_{n}} A_{a p_{1} p_{2} \cdots p_{n}, b q_{1} q_{2}}^{*}$ as an example, the Feynman propagator only considers the temporal correlation, and bringing the expression of the Feynman propagator into this formula, we can get
\begin{equation}\label{equa2}
\begin{aligned}
&A_{a_{1} i_{2}-i_{2}, b_{1} j_{2}-j_{n}} A_{a_{2} p_{2}-p_{n}, b_{4}, q_{2}-q_{n}} \\
=&\iint_{\omega_{0}-\frac{1}{2}\Delta \omega}^{\omega_{0}+\frac{1}{2} \Delta \omega} f\left(\omega_{a}\right) e^{-i\left[\omega_{a}\left(i_{1}+i_{2} \cdots+i_{n}-p_{1}-p_{2} \cdots-p_{n}\right)\right]}\\ & \cdot f\left(\omega_{b}\right) e^{-i\left[\omega_{b}\left(j_{1}+j_{2} \cdots+j_{n}-q_{1}-q_{2} \cdots-q_{n}\right)\right]} d \omega_{a} d \omega_{b}  ,
\end{aligned}
\end{equation}
where we define that $i_{1}, i_{2}, \cdots i_{n}, j_{1}, j_{2} \cdots j_{n}=t_{1}, t_{2}$, $p_{1}, p_{2} \cdots p_{n}, q_{1}, q_{2} \cdots q_{n}=t_{1}, t_{2}$, and they represent the time that each photon reaches channel 1 and channel 2 when it passes through a MI. In order to make Eq. (\ref{equa2}) be a constant term, it must also satisfy
\begin{equation}\left\{\begin{array}{l}\label{equa3}
i_{1}+i_{2} \cdots+i_{n}-p_{1}-p_{2} \cdots-p_{n}=0 \\
j_{1}+j_{2} \cdots+j_{n}-q_{1}-q_{2} \cdots-q_{n}=0.
\end{array}\right .
\end{equation}

Here we can use the idea of permutation and combination to solve this equation. Taking photon $a$ as an example:

1. When there is no $t_{1}$ in $i_{1}, i_{2}, \cdots i_{n}$, the number of combinations contained is $C_{N}^{0}$, and at the same time all of the $t_{1}$ in $p_{1}, p_{2}, \cdots p_{n}$ must be 0 to make $i_{1}+i_{2} \cdots+i_{n}-p_{1}-p_{2} \cdots-p_{n}=0$ possible, then the total number of combinations is $\left(C_{N}^{0}\right)^{2}$.

2. When there is one $t_{1}$ in $i_{1}, i_{2}, \cdots i_{n}$, the number of combinations contained is $C_{N}^{1}$. And we also have to have one $t_{1}$ in  $p_{1}, p_{2}, \cdots p_{n}$ that are equal to $C_{N}^{1}$, then the total number of combinations is  $\left(C_{N}^{1}\right)^{2}$.

3. When there are two  $t_{1}$ in  $i_{1}, i_{2}, \cdots i_{n}$, the number of combinations contained is  $C_{N}^{1}$. and at the same time there are also have two  $t_{1}$ in  $p_{1}, p_{2}, \cdots p_{n}$, then the total number of combinations is  $\left(C_{N}^{2}\right)^{2}$.

4. By analogy, when there are N   $t_{1}$in  $i_{1}, i_{2}, \cdots i_{n}$, the number of combinations included is $C_{N}^{N}$. In the meantime, and it must also satisfy all N $t_{1}$ in $p_{1}, p_{2}, \cdots p_{n}$, finally the total number of combinations at this time is  $\left(C_{N}^{N}\right)^{2}$.

At this time, we add up all the above combination numbers, it can be expressed as
\begin{equation}\label{equa4}
\left(C_{N}^{0}\right)^{2}+\left(C_{N}^{1}\right)^{2}+\left(C_{N}^{2}\right)^{2}+\cdots+\left(C_{N}^{N}\right)^{2}
=\sum_{n=0}^{N}\left(C_{N}^{n}\right)^{2}.
\end{equation}
We just discussed the case of photon $a$ above, meanwhile the case of photon $b$ is the same and needs to be satisfied  $i_{1}+i_{2} \cdots+i_{n}-p_{1}-p_{2} \cdots-p_{n}=0$. Therefore, we end up with the coefficient of the constant term as
\begin{equation}\label{equa5}
C_{c}=\left(\sum_{n=0}^{N}\left(C_{N}^{n}\right)^{2}\right)^{2}.
\end{equation}

\section{The coefficient of the second-order coherence terms $C_{i}$ of chaotic light transferring through the Michelson interferometer $N$ times}\label{Appendix B}
What following is the derivation of the coefficients of the second-order coherence terms. Taking the photon $a$ as an example, through the study of the second-order coherence terms, we find that when Eq. (\ref{equa2}) satisfies
\begin{equation}\label{equa6}
i_{1}+i_{2} \cdots+i_{n}-p_{1}-p_{2} \cdots-p_{n}=k\left(t_{1}-t_{2}\right),
\end{equation}
where $k=-N,-(N-1), \cdots-1,1, \cdots(N-1), N$. In order to simplify our derivation process, we first consider when $k \geq 1$, that is $k=1, \cdots(N-1), N$, this kind of calculation result is half of all the results, the discussions are as below:

1. When  $i_{1}+i_{2} \cdots+i_{n}-p_{1}-p_{2} \cdots-p_{n}=\left(t_{1}-t_{2}\right)$, there are $N$ cases in total:

1.1. When there is one $t_{1}$ in $i_{1}, i_{2}, \cdots i_{n}$, the number of combinations contained is  $C_{N}^{1}$, and at the same time all of the $t_{1}$ in $p_{1}, p_{2}, \cdots p_{n}$ must be 0 to make  $i_{1}+i_{2} \cdots+i_{n}-p_{1}-p_{2} \cdots-p_{n}=\left(t_{1}-t_{2}\right)$ possible, then the total number of combinations is $C_{N}^{1} C_{N}^{0}$ .

1.2. When there is two  $t_{1}$ in  $i_{1}, i_{2}, \cdots i_{n}$, the number of combinations contained is $C_{N}^{2}$. And we also have to have one $t_{1}$ in $p_{1}, p_{2}, \cdots p_{n}$ that are equal to $C_{N}^{1}$, then the total number of combinations is  $C_{N}^{2} C_{N}^{1}$.

1.3. When there are three $t_{1}$ in $i_{1}, i_{2}, \cdots i_{n}$, the number of combinations contained is $C_{N}^{3}$. And at the same time there are also two $t_{1}$ in $p_{1}, p_{2}, \cdots p_{n}$, finally the total number of combinations is  $C_{N}^{3} C_{N}^{2}$.

1.4. By analogy, when there are N  $t_{1}$ in  $i_{1}, i_{2}, \cdots i_{n}$, the number of combinations included is  $C_{N}^{N}$, and all of them must satisfy $(N-1) \cdot t_{1}$, then the whole number of combinations at this time is  $C_{N}^{N} C_{N}^{N-1}$.

In summary, the result of the addition of all the above cases just considers the case of photon $a$, and also the photon $b$ situation is exactly the same. Meanwhile when we take the value of $k$, all the combinations can be expressed at this time for:
\begin{equation}\label{equa7}
\begin{aligned} C_{1} &=2 \cdot\left(C_{N}^{1} C_{N}^{0}+C_{N}^{2} C_{N}^{1}+C_{N}^{3} C_{N}^{2}+\cdots+C_{N}^{N} C_{N}^{N-1}\right)^{2} \\ &=2 \left(\sum_{n=0}^{N-1} C_{N}^{n} C_{N}^{n+1}\right)^{2}. \end{aligned}
\end{equation}

2. When $i_{1}+i_{2} \cdots+i_{n}-p_{1}-p_{2} \cdots-p_{n}=2 \cdot\left(t_{1}-t_{2}\right)$, there are $\left(N-1\right)$ cases in total:

2.1. When there is two $t_{1}$ in $i_{1}, i_{2}, \cdots i_{n}$, the number of combinations contained is $C_{N}^{2}$. And at the same time all of the $t_{1}$ in $p_{1}, p_{2}, \cdots p_{n}$ must be 0 to make  $i_{1}+i_{2} \cdots+i_{n}-p_{1}-p_{2} \cdots-p_{n}=2 \cdot\left(t_{1}-t_{2}\right)$ possible, then the total number of combinations is  $C_{N}^{2} C_{N}^{0}$.

2.2. When there is three  $t_{1}$ in  $i_{1}, i_{2}, \cdots i_{n}$, the number of combinations contained is  $C_{N}^{3}$. And we also have one $t_{1}$ in $p_{1}, p_{2}, \cdots p_{n}$ that are equal to $t_{1}$, then the total number of combinations is $C_{N}^{3} C_{N}^{1}$.

2.3. When there are four $t_{1}$ in $i_{1}, i_{2}, \cdots i_{n}$, the number of combinations contained is $C_{N}^{4}$. And at the same time there are also two $t_{1}$ in $p_{1}, p_{2}, \cdots p_{n}$, finally the total number of combinations is $C_{N}^{4} C_{N}^{2}$.

2.4. By analogy, when there are N $t_{1}$ in $i_{1}, i_{2}, \cdots i_{n}$, the number of combinations included is $C_{N}^{N}$, and at the same time all must satisfy  $(N-2) \cdot t_{1}$ , then the number of combinations is  $C_{N}^{N} C_{N}^{N-2}$.

In this way, the result of the addition of all the above cases only considers the case of photon $a$, and the case of photon $b$ is the same. Meanwhile, when we take the value of $k$ as $k=-N,-(N-1), \cdots-1,1, \cdots(N-1), N$, the number of combinations can be expressed as
\begin{equation}\label{equa8}
\begin{aligned} C_{2} &=2 \cdot\left(C_{N}^{2} C_{N}^{0}+C_{N}^{3} C_{N}^{1}+C_{N}^{4} C_{N}^{2}+\cdots+C_{N}^{N} C_{N}^{N-2}\right)^{2} \\ &=2 \left(\sum_{n=0}^{N-2} C_{N}^{n} C_{N}^{n+2}\right)^{2}. \end{aligned}
\end{equation}

3. Similarly to the situation discussed above, we can successively express $C_{1}, C_{2}, C_{3} \cdots C_{N}$. When $i_{1}+i_{2} \cdots+i_{n}-p_{1}-p_{2} \cdots-p_{n}=N \cdot\left(t_{1}-t_{2}\right)$, there is only one case that have N  $t_{1}$ in  $i_{1}, i_{2}, \cdots i_{n}$, the number of combinations included was $C_{N}^{N}$. And at the same time, $p_{1}, p_{2}, \cdots p_{n}$ must meet that have no $t_{1}$, then the number of combinations is  $C_{N}^{N} C_{N}^{0}$. When we consider the case of photons $b$ and the value range of $K$, the total number of combinations at this time is expressed as
\begin{equation}\label{equa9}
C_{N}=2 \cdot\left(C_{N}^{N} C_{N}^{0}\right)^{2}.
\end{equation}

To sum up, all the discussions are added up as follows
\begin{equation}\label{equa10}
\begin{aligned}
C_{i} &=C_{1}+C_{2}+\cdots C_{N} \\&=\sum_{m=1}^{N} 2 \left(\sum_{n=0}^{N-m} C_{N}^{n} C_{N}^{m+n}\right)^{2}. \end{aligned}
\end{equation}

\bibliographystyle{apsrev4-2}
\bibliography{apssamp}

\end{document}